\documentclass[a4paper]{article}

\usepackage{INTERSPEECH2022}

\usepackage{amsfonts,amssymb} 
\usepackage{bbding}  
\usepackage{caption} 
\usepackage{tabularx} 
\usepackage{float}
\usepackage{array}  
\usepackage{booktabs} 
\usepackage{multirow}
\usepackage{makecell} 
\usepackage{color}

\title{CT-SAT: Contextual Transformer for Sequential Audio Tagging}
\name{Yuanbo Hou$^1$, Zhaoyi Liu$^2$, Bo Kang$^1$, Yun Wang$^3$, Dick Botteldooren$^1$}
\address{
  $^1$Ghent University, Belgium  \qquad $^2$KU Leuven, Belgium \qquad  $^3$Meta AI, USA }
\email{\{yuanbo.hou, bo.kang, Dick.Botteldooren\}@UGent.be \\
zhaoyi.liu@student.kuleuven.be, maigoakisame@gmail.com}

\begin{document}

\maketitle
\begin{abstract}
Sequential audio event tagging can provide not only the type information of audio events, but also the order information between events and the number of events that occur in an audio clip.
Most previous works on audio event sequence analysis rely on connectionist temporal classification (CTC).  
However, CTC's conditional independence assumption prevents it from effectively learning correlations between diverse audio events.
This paper first attempts to introduce Transformer into sequential audio tagging, since Transformers perform well in sequence-related tasks.
To better utilize contextual information of audio event sequences, we draw on the idea of bidirectional recurrent neural networks, and propose a contextual Transformer (cTransformer) with a bidirectional decoder that could exploit the forward and backward information of event sequences.
Experiments on the real-life polyphonic audio dataset show that, compared to CTC-based methods, the cTransformer can effectively combine the fine-grained acoustic representations from the encoder and coarse-grained audio event cues to exploit contextual information to successfully recognize and predict audio event sequences.
 
\end{abstract}
\vspace{-0.2cm}
\section{Introduction}
Audio Tagging (AT) is a multi-label classification task that identifies which target audio events occur in an audio clip. 
AT only predicts the type of events occurring in an audio clip, not the order between these events nor how many times they occur.
Audio events naturally occur sequentially in a sequence, and there is often a relationship between the preceding and following events.  
This paper studies sequential audio tagging (SAT), which aims to learn such relationships between events and predict sequences of audio events in audio clips.
SAT can be applied for tasks such as audio classification \cite{audio_cl}, audio captioning \cite{audio_cap}, acoustic scene analysis \cite{acoustic_scene}, and event anticipation \cite{event_pred}.


Previous works related to SAT mostly rely on connectionist temporal classification (CTC) \cite{ctc} to identify event sequences.
Paper \cite{dcase2018_ctc} explores the possibility of polyphonic SAT using sequential labels and utilizes CTC to train convolutional recurrent neural networks (CRNN) \cite{crnn} with learnable gated linear units (GLU) \cite{GLU} to tag event sequences.
As audio events often overlap with each other, 
the order of start and end boundaries of events is used in \cite{dcase2018_ctc} as sequential labels. 
For example, the double-boundary sequential label of an audio clip 
might be 
\textit{\footnotesize dishes\_start, dishes\_end, speech\_start, blender\_start, speech\_end, speech\_start, blender\_end, speech\_end}.
Sequential labels do not contain the onset and offset time information of audio events, which avoids the problem of inaccurate 
annotations of frame-level labels, and reduces the annotation workload. 
 In addition to exploring the feasibility of recognizing audio event sequences in SAT, 
 CTC-based methods have also been attempted for sound event detection (SED),
 which detects the type,
starting time, and ending time of audio events.
A bidirectional long short-term memory (LSTM) RNN \cite{lstm} equipped with CTC (BLSTM-CTC) \cite{wangyun} is used to detect events using double-boundary sequential labels.
The results \cite{wangyun} on a very noisy corpus show that BLSTM-CTC is able to locate boundaries of 
audio events with rough hints about their positions.
Apart from methods using double-boundary labels, 
another CTC-based SED system \cite{hou2019sound} uses single-boundary sequential labels (the order of start boundary of events) with unsupervised clustering to detect the type and occurrence time of audio events.
 CTC redefines the loss function of RNN \cite{ctc} and allows it to be trained for 
 sequence-related
 tasks to keep the order information of events. 
 However, CTC implicitly assumes that outputs of the network at different time steps are conditionally independent \cite{ctc}, which makes CTC-based approaches unable to effectively learn the contextual information inherent in audio event sequences.
This paper attempts to introduce Transformers~\cite{Transformer}, which have revolutionized the field of natural language processing~\cite{nlp}, into SAT.
Transformer \cite{Transformer} does not have the conditional independence assumption in CTC.
 Compared with RNN-based models, Transformer can access information at any time step from any other time step, thereby capturing long-term dependencies~\cite{long_term} between audio events. 
In addition, the training of Transformer can also be efficiently parallelized.


\label{ssec:figure-f}
\begin{figure*}[ht]
	\setlength{\abovecaptionskip}{0.1cm}  
	\setlength{\belowcaptionskip}{-0.55cm}   
	\centerline{\includegraphics[width = 0.86 \textwidth]{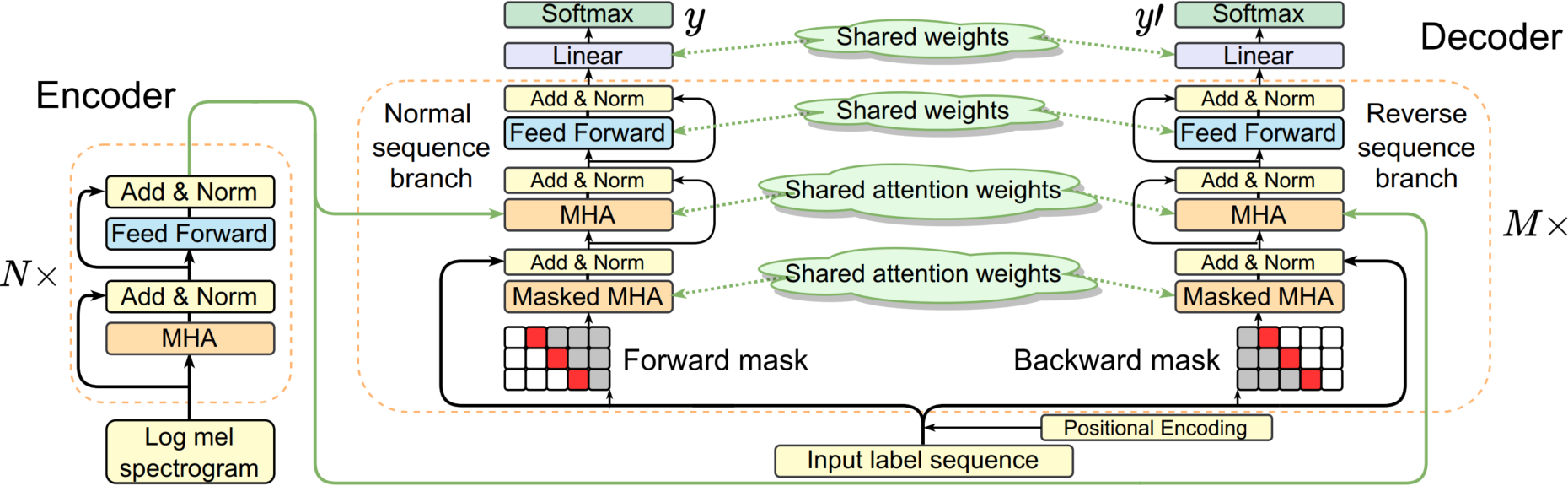}}
	\caption{The proposed contextual Transformer. In the forward and backward mask, the red, gray, and white blocks indicate the masked position of the information to be predicted, the position of the masked information, and the position of the available information.}
	\label{model}
\end{figure*}

\vspace{-0.05cm}
When learning sequence information, the decoder in Transformer \cite{Transformer} exploits past information to infer the upcoming event.
For example, when recognizing audio event sequences 
\textit{\footnotesize ``fire, alarm, run''}
and 
\textit{\footnotesize ``fire, crying, sobbing''},
the model may be confused between \textit{\footnotesize alarm} and \textit{\footnotesize crying}
when forward inferring the next event from 
\textit{\footnotesize fire}.
But if the target event is backward inferred from 
\textit{\footnotesize run} and \textit{\footnotesize sobbing}
respectively, the probability of 
\textit{\footnotesize alarm} and \textit{\footnotesize crying}
is different in different sequences.
Contextual information can help the model learn the differences between sequences in detail. 
To make more comprehensive utilize the contextual information in audio event sequences, this paper draws on the idea of bidirectional RNN \cite{brnn} and proposes a contextual Transformer (cTransformer) to explore the bidirectional information of audio event sequences.
The cTransformer consists of the encoder and decoder, the latter of which is the main contribution of this paper.
The decoder attempts to fuse frame-level representations from the encoder with the clip-level event cues to infer the target by 
combing the forward and backward information learned from normal and reverse sequences, respectively. 
Then, the loss between the prediction from normal sequence branch and the prediction from reverse sequence branch is calculated and fed back to update parameters to learn a more accurate prediction about the same target. 
During training, partial weights of the normal and reverse sequence branches are shared, 
these shared weights can learn both the forward and backward information.
That is, with the help of shared weights, the decoder is able to learn contextual information simultaneously to more comprehensively and accurately identify audio event sequences.

The contributions of this paper are: 
1) this paper introduces Transformer into SAT;
2) The cTransformer that can utilize bidirectional information is proposed to better identify audio event sequences in audio clips;
3) To explore the feasibility of SAT based on cTransformer, we manually label sequential labels for a polyphonic audio dataset from real life, and compare the performance of cTransformer and other CTC-based methods on it.
This paper is organized as follows, Section 2 shows cTransformer. Section 3 describes the dataset, experimental setup, and analyzes the results. Section 4 gives conclusions.


\vspace{-0.3cm}
\section{Contextual Transformer}

Motivated by the performance of Transformer in sequence modeling \cite{Transformer, devlin2018bert}, 
and the significance of contextual information in audio tasks \cite{contextual_audio, contextual_audio2, contextual_audio3}, this paper proposes cTransformer for audio event sequence analysis. 
The cTransformer aims to transform the acoustic feature to the corresponding event sequential label using both global information and rich contextual
details. 


\vspace{-0.25cm}
\subsection{Data preparation} 


In audio event tasks, the most frequently used acoustic feature is the log mel spectrogram \cite{mel}.
The audio clip $x$ is converted to the time-frequency representation $X(t; f)$ of log mel spectrogram and input to the model. 
Referring to \cite{hou2019sound}, the start boundary order of
events is used as sequential labels.
For the normal sequence branch 
in Figure~\ref{model},
given the sequential label $y$ is 
\textit{\footnotesize \textless{S}\textgreater, $event_1$, $event_2$, ..., $event_k$, \textless{E}\textgreater},
where $k$ means the $k$-th event, 
\textit{\footnotesize \textless{S}\textgreater} and \textit{\footnotesize \textless{E}\textgreater}
are the default tokens \cite{Transformer} indicating the start and end of prediction, respectively.
For the reverse sequence branch,
the sequential label $y\prime$ is 
\textit{\footnotesize $\textless{S'}\textgreater$, $event_k$, $event_{k-1}$, ..., $event_1$, \textless{E}\textgreater},
where \textit{\footnotesize $\textless{S'}\textgreater$} is the token indicating the start of reverse sequence prediction.
The sequential label $y$ of 
an audio clip might be 
\textit{\footnotesize ``\textless{S}\textgreater, dishes, speech, speech, blender, speech, \textless{E}\textgreater''}.
The corresponding $y\prime$ is 
\textit{\footnotesize ``$\textless{S'}\textgreater$, speech, blender, speech, speech, dishes, \textless{E}\textgreater''}.
 


\vspace{-0.25cm}
\subsection{Encoder in contextual Transformer}


The encoder aims to convert input acoustic features into high-level representations.
To consider the audio information globally, this paper does not divide input features into small patches \cite{ast}, so there is no positional encoding \cite{Transformer} in the encoder.
The encoder mainly consists of $N$ identical blocks with multi-head attention layers (MHA) and feed forward layers, which are analogous to the encoder in Transformer \cite{Transformer}.
The attention function in MHA is scaled dot-product attention, whose input consists of
queries and keys of dimension $d_{k}$, and values of dimension $d_{v}$ \cite{Transformer}. 
The attention is calculated on a set of queries, keys, and values packed into matrix $\mathbf{Q}$, $\mathbf{K}$, and $\mathbf{V}$, respectively.
\begin{equation}
\setlength{\abovedisplayskip}{1pt}
\setlength{\belowdisplayskip}{2pt}
Attention(\mathbf{Q, K, V} )=softmax(\mathbf{QK^T} / \sqrt{d_{k} } )\mathbf{V} 
\label{self-attention}
\end{equation}
Then, MHA is used to allow the model to jointly focus on representations from different subspaces at different positions. \begin{equation}
\setlength{\abovedisplayskip}{1pt}
\setlength{\belowdisplayskip}{2pt}
\begin{split}
MHA(\mathbf{Q,K,V})=Concat(head_{1}, ..., head_{h} )\mathbf{W}^{O} 
\\
where \quad head_{i}= Attention(\mathbf{QW}_{i}^{Q}, \mathbf{KW}_{i}^{K}, \mathbf{VW}_{i}^{V}) 
\end{split}
\end{equation}
Where $head_{i}$ represents the output of the $i$-th attention head for a total number of $h$ heads. 
$\mathbf{W}_{i}^{Q}$, $\mathbf{W}_{i}^{K}$, $\mathbf{W}_{i}^{V}$ and $\mathbf{W}^{O}$ are learnable weights.
For MHA in the encoder, 
$\mathbf{Q}$, $\mathbf{K}$, and $\mathbf{V}$ come from the same place, at this point, the attention in MHA is called self-attention \cite{Transformer}.
Next, the feed forward layer that consists of two linear transformations with ReLU activation function \cite{ReLU} in between is applied.
For the parameters involved above, all refer to the default settings of Transformer \cite{Transformer}.

\vspace{-0.25cm}
\subsection{Decoder in contextual Transformer}


The cTransformer is expected to efficiently capture contextual information in audio event sequences without reducing the Transformer's global summarization ability. 
The global attention in the encoder can attend to the information of each position.
However,
the self-attention in Masked MHA of decoder relies only on the forward information to sequentially predict the next event 
to preserve autoregressive property \cite{Transformer}, as the normal sequence branch in Figure~\ref{model}.
Thus, a bidirectional sequence decoder that can exploit the forward and backward information is proposed, as shown in the decoder of Figure~\ref{model}.
To enhance the ability of the model to capture the contextual information of the target event, 
the normal and reverse sequence branches jointly predict the same target each time.
Since some weights of the two branches are shared, these 
weights both learn forward information about the target and capture its related backward information to help the model learn and model the contextual information about each target more accurately.

The decoder consists of two branches with the same structure, and each branch contains $M$ identical blocks,
which are analogous to the decoder in Transformer \cite{Transformer}.
In Masked MHA, the forward and backward masks are to block future information and past information to preserve the autoregressive property, respectively.
Positions corresponding to invisible information will be masked with $-\infty$ \cite{Transformer}.
The attention in Masked MHA is self-attention, which means that $\mathbf{Q}$, $\mathbf{K}$, and $\mathbf{V}$ all come from the input sequence of event labels. 
In the next MHA used to fuse frame-level acoustic representations from the encoder and clip-level event cues, for the encoder-decoder attention, $\mathbf{Q}$ is from the previous decoder layer, while $\mathbf{K}$ and $\mathbf{V}$ are from the output of the encoder.
For the $t$-th target $event_t$, 
given the input embedding for the normal sequence branch is $\overrightarrow{\mathbf{z}}_{t-1}$, and the input embedding for the reverse sequence branch is $\overleftarrow{\mathbf{z}}_{t+1}$.
Let $\overrightarrow{\mathbf{p}}_{t}$ and  $\overleftarrow{\mathbf{p}}_{t}$ be the prediction for $event_t$ from the normal and reverse sequence branches.
For the normal sequence branch exploring forward information, $\overrightarrow{\mathbf{p}}_{t}$ is jointly derived from 
the output of encoder $\mathbf{O}_{En}$ and embedding $\overrightarrow{\mathbf{z}}_{t-1}$ after forward Masked MHA $M_{f}$.
For the reverse sequence branch exploring backward information, $\overleftarrow{\mathbf{p}}_{t}$ is jointly derived from $\mathbf{O}_{En}$ and embedding $\overleftarrow{\mathbf{z}}_{t+1}$ after backward Masked MHA $M_{b}$.
\begin{equation}
\setlength{\abovedisplayskip}{2pt}
\setlength{\belowdisplayskip}{2pt}
\begin{split}
\overrightarrow{\mathbf{p}}_{t} =\phi (M_{f}(\overrightarrow{\mathbf{z}}_{t-1})+\mathbf{O}_{En}\mathbf{W}_{MHA}^{(f)}+\mathbf{b}^{(f)})
\\
\overleftarrow{\mathbf{p}}_{t} =\phi (M_{b}(\overleftarrow{\mathbf{z}}_{t+1})+\mathbf{O}_{En}\mathbf{W}_{MHA}^{(b)}+\mathbf{b}^{(b)}) 
\end{split}
\end{equation}
where 
$\mathbf{b}^{(f)}$ and $\mathbf{b}^{(b)}$ are biases in  normal and reverse sequence branches, 
$\mathbf{W}_{MHA}^{(f)}$ and $\mathbf{W}_{MHA}^{(b)}$ are learnable weights in MHA, $\phi$ denote the set of mapping functions in each branch of the decoder. 
In the inference phase, the model uses the normal sequence branch for prediction.
The remaining layers and parameters in the decoder are the same as those of Transformer \cite{Transformer}.

\vspace{-0.2cm}
\subsection{Loss function in contextual Transformer}

Denote $\overrightarrow{\mathbf{p}}_{t}$ and  $\overleftarrow{\mathbf{p}}_{t}$ as $p$ and $p\prime$, and the corresponding ground-truth labels are $y$ and $y\prime$, respectively.
Following the loss function in Transformer \cite{Transformer}, 
cross entropy (CE) loss is used as the loss function for the normal and reverse sequence branch to compute the normal and reverser sequential tagging loss.
\begin{equation}
\setlength{\abovedisplayskip}{2pt}
\setlength{\belowdisplayskip}{2pt}
\mathcal{L}_{normal} = CE(p, y), 
\quad
\mathcal{L}_{reverse}=  CE(p\prime, y\prime)
\label{ce}
\end{equation}
Since $p$ and $p\prime$ are the prediction for the same target, 
the mean squared error (MSE) loss that performs well
in regression tasks \cite{mse}\cite{mse-regression}\cite{mse-classification} is used 
as the context loss 
to measure the distance between $p$ and $p\prime$ in the latent space.
\begin{equation}
\setlength{\abovedisplayskip}{2pt}
\setlength{\belowdisplayskip}{2pt} 
\mathcal{L}_{context} = MSE(p\prime, p)
\label{mse}
\end{equation}
To consider the forward and backward information at the same time in training phase, losses of different branches are calculated together. The final loss of the cTransformer is
\begin{equation}
\setlength{\abovedisplayskip}{2pt}
\setlength{\belowdisplayskip}{2pt}
\mathcal{L} = \lambda_{n}\mathcal{L}_{normal}+\lambda_{r}\mathcal{L}_{reverse}+\lambda_{c}\mathcal{L}_{context} 
\label{loss}
\end{equation}
where $\lambda$ 
adjusts the weights of different loss components during training. 
$\lambda$ defaults to 1.
During the training process, the forward prediction $p$ and backward prediction $p\prime$ will be aligned to capture the rich contextual information around the target event and learn the entire sequence embeddings more accurately.

\begin{table}[b]\footnotesize
	\setlength{\abovecaptionskip}{0cm}   
	\setlength{\belowcaptionskip}{-0.47cm}   
	\renewcommand\tabcolsep{1.5pt} 
	\centering
	\caption{Results of the model with different ratios of $N$ and $M$.}
	\begin{tabular}
	{p{0.5cm}<{\centering}|
	p{1.2cm}<{\centering}|
	p{0.9cm}<{\centering}
	p{0.9cm}<{\centering}|
	p{0.5cm}<{\centering}|
	p{1.2cm}<{\centering}|
	p{0.9cm}<{\centering}
	p{0.9cm}<{\centering}} 

\hline

		{\#} & {\{$N$, $M$\}} & {\textsl{AUC}} & {\textsl{BLEU}} & {\#} &  {\{$N$, $M$\}} & {\textsl{AUC}} & {\textsl{BLEU}} \\ 
		
		\specialrule{0em}{0.1pt}{0.1pt}
		
        \hline
        \specialrule{0em}{0.1pt}{0.1pt}
		 
		1 & {\{1, 1\}} &  0.771  &  0.468 & 7 & {\{3, 3\}} &  0.784  &  0.482  \\
		
		2 & {\{1, 2\}} &  \textbf{0.800 }  & \textbf{0.491}  & 8 & {\{3, 6\}} &  0.770  &  0.472  \\
		
		3 & {\{2, 2\}} &  0.775   & 0.481  & 9 & {\{4, 2\}} &  0.779  &  0.467 \\
		
		4 & {\{2, 4\}} &  0.775  &  0.483  & 10 & {\{4, 4\}} &  0.787  &  0.464  \\
		
		5 & {\{2, 5\}} &  0.783  &  0.473 & 11 & {\{5, 5\}} &  0.774  & 0.461  \\
		
		6 & {\{3, 1\}} &  0.782  &  0.474 & 12 & {\{6, 6\}} &  0.778  &  0.456 \\
		 
		\hline
	\end{tabular}
	\label{tab:blocks}
\end{table}

\vspace{-0.2cm}
\section{Experiments and results}

\vspace{-0.1cm}
\subsection{Dataset, Baseline, Experiments Setup, and Metrics}
Since there is no publicly available polyphonic audio dataset with sequential labels, we manually label the DCASE domestic environment audio dataset \cite{dcase2018} with the start boundary order of events as sequential labels by referring to \cite{hou2019sound}, and release the sequential label set to motivate more relevant research.
The domestic audio dataset excerpted from Audioset \cite{aduioset} contains 10 classes of real-life polyphonic audio events, where the training and test sets consist of 1578 and 288 audio clips, respectively. During training, the validation set is randomly composed of 20\% of the samples in the training set.
After manual annotation and cross-checking, the number of occurrence events contained in the train set and test set is 3619 and 923, where the length of the longest audio event sequences is 20 and 14, respectively.

Most previous audio event sequence analysis works rely on CTC, so BLSTM-CTC \cite{wangyun} is used as Baseline. 
This paper also compares the cTransformer with CTC-based convolutional bidirectional gated recurrent units (CBGRU-CTC) \cite{csps_ctc}, and CBGRU-CTC equipped with GLU in convolutional layers (CGLU-BGRU-CTC) \cite{dcase2018_ctc}, and in both convolutional and recurrent layers (CBGRU-GLU-CTC) \cite{hou2019sound}.

In training, log mel-band energy with 64 banks \cite{mel} is extracted using STFT with Hamming window length of 46 ms and the overlap
is $1/3$ between windows following the settings of \cite{dcase_kong}.
Stochastic gradient descent with momentum (SGDM) \cite{sgdm} with an initial learning rate of 1e-3,  batch size of 64, and momentum value of 0.9 is used to minimize the loss.
Dropout \cite{dropout} and layer normalization \cite{layernorm} are used to prevent over-fitting. 
Systems are trained on a single card Tesla V100-SXM2-32GB for maximum 1000 epochs.
For more details, source code, and the manually labeled dataset with sequential labels, 
please visit the project homepage
(\textcolor{blue}{\underline{https://github.com/Yuanbo2020/Contextual-Transformer}}).


SAT consists of AT plus order information between events.
This paper uses precision (\textsl{P}), recall (\textsl{R}),  F-score (\textsl{F}), accuracy (\textsl{Acc}) \cite{metrics}, and area under curve (\textsl{AUC}) \cite{AUC} to measure the results of AT in various aspects to show the performance of models on the basic event recognition. 
Then, the bilingual evaluation understudy (\textsl{BLEU}) \cite{bleu} commonly used in sequence tasks is adopted to comprehensively evaluate the SAT results. 
Higher \textsl{P}, \textsl{R}, 
\textsl{F},
\textsl{Acc}, \textsl{AUC}, and \textsl{BLEU} indicate a better performance.

\vspace{-0.2cm}
\subsection{Results and Analysis}

The encoder and decoder of the cTransformer consist of $N$ and $M$ identical blocks, respectively. 
This paper first explores the optimal ratio of blocks of encoder and decoder to determine the final model structure, as shown in Table \ref{tab:blocks}.
SAT is equivalent to AT with additional sequence information of events. 
So, \textsl{AUC}, which can avoid the influence of different threshold interference, is used to measure the results of AT more comprehensively, and \textsl{BLEU} is used to evaluate the results of SAT.

In Table \ref{tab:blocks}, the performance of 
the model
does not increase monotonically with the number of blocks. 
When \{$N$, $M$\} is \{1, 2\}, the model achieves the best results on the test dataset. 
In Transformer \cite{Transformer}, \{$N$, $M$\} defaults to \{6, 6\}. 
The size of the best model in this paper is smaller than that of Transformer, and the reason may be that the polyphonic audio dataset with manually labeled sequential labels 
is not large scale, resulting in smaller models with fewer blocks performing well.
And in the experiment, we found that the model will show more serious overfitting when the values of $N$ and $M$ are large.

\begin{table}[b]\footnotesize   
	\setlength{\abovecaptionskip}{0cm}   
	\setlength{\belowcaptionskip}{-0.47cm}  
	\renewcommand\tabcolsep{1pt} 
	\centering
	\caption{Ablation experiments of the cTransformer on test set.}
	\begin{tabular}
	{p{0.4cm}<{\centering}|
	p{1.2cm}<{\centering}
	p{1.2cm}<{\centering}
	p{1.2cm}<{\centering}|
	p{0.8cm}<{\centering}
	p{0.97cm}<{\centering}
	p{0.7cm}<{\centering}
	p{1cm}<{\centering}}
	
    \hline
	   {\#} & $\mathcal{L}_{normal}$
	   & $\mathcal{L}_{reverse}$ 
	   & $\mathcal{L}_{context}$
	   & \textsl{F (\%)} & \textsl{Acc (\%)}
	   & \textsl{AUC} & \textsl{BLEU}\\
		\hline  
		\specialrule{0em}{0.05em}{0.pt}
		1 & \CheckmarkBold & \XSolidBrush  & \XSolidBrush  &  66.42 &  90.41 & 0.780 & 0.474 \\ 
		2 &  \XSolidBrush & \CheckmarkBold  & \XSolidBrush  &  64.58 &  89.79 & 0.765 & 0.472 \\
		3 &  \CheckmarkBold & \CheckmarkBold  & \XSolidBrush  &  67.39 &  90.66 & 0.785 & 0.489 \\
		4 &  \CheckmarkBold & \CheckmarkBold  & \CheckmarkBold  & \textbf{70.42} &  \textbf{91.63} & \textbf{0.800} & \textbf{0.491} \\
	\hline
	\end{tabular}
	\label{tab:ablation}
\end{table}

The next step is to optimize the scaling factor $\lambda$ of different losses. 
Different losses target different information.
The $\mathcal{L}_{context}$ with MSE aims to align predictions of the normal and reverse sequence branches to make their predictions of the current event more consistent, 
while $\mathcal{L}_{normal}$ and $\mathcal{L}_{reverse}$ focus on learning task-goal-oriented representations to improve the accuracy of individual event sequence recognition. 
Table \ref{tab:ablation} conducts ablation studies to imply the importance of the information represented by different losses to the cTransformer.

\begin{table}[b]\footnotesize 
	\setlength{\abovecaptionskip}{0cm}   
	\setlength{\belowcaptionskip}{-0.4cm}   
	\renewcommand\tabcolsep{1.5pt} 
	\centering 
\caption{The effect of different $\lambda$ values on the cTransformer.}
	\begin{tabular}
	{
	p{0.3cm}<{\centering}|
	p{0.5cm}<{\centering}
	p{0.5cm}<{\centering}
	p{0.5cm}<{\centering}|
	p{0.75cm}<{\centering}
	p{0.85cm}<{\centering}|
	p{0.3cm}<{\centering}|
	p{0.5cm}<{\centering}
	p{0.5cm}<{\centering}
	p{0.5cm}<{\centering}|
	p{0.75cm}<{\centering}
	p{0.85cm}<{\centering}
	}  
		\hline 
		{\#} & 
		$\lambda_{n}$ &  $\lambda_{r}$ & $\lambda_{c}$ 
		& \textsl{AUC}  
		& \textsl{BLEU} 
		& {\#} & 
		$\lambda_{n}$ &  $\lambda_{r}$ & $\lambda_{c}$ 
		& \textsl{AUC}  
		& \textsl{BLEU}\\ 
        \hline
     
     1 & 1  & 0.5  & 0.1 &  0.789  & 0.481  
     & 8 & 0.5 &  1  & 1 &  0.774  &   0.467 \\
     
     2 & 1  & 0.5  & 0.25 &  0.803  &  \textbf{0.511}
     & 9 & 1 &  1  & 0.1 &  0.791  &  0.485 \\ 
     
     3 & 1  & 0.5  & 0.5 &   0.782  &   0.488
     & 10 & 1 &  1 &  0.25 &  0.788   &  0.501 \\ 
     
     4 & 1  &  0.5 &  1 &  \textbf{0.805}  &  0.505  
     & 11 & 1 &  1 &  0.5 &  0.783  & 0.487 \\ 
     
     5 & 0.5 &  1  & 0.1 &  0.784  &   0.479
     & 12 & 0.1 &  0.1 &  1 &  0.763  & 0.465 \\ 
     
     6 & 0.5 &  1  & 0.25 &  0.788 &  0.482
     & 13 & 0.25 &  0.25 & 1 &  0.774  &  0.466 \\ 
     
     7 & 0.5  & 1  & 0.5 &  0.778  &   0.465
      & 14 & 0.5 & 0.5  & 1 &  0.785  &  0.472 \\ 
     
\hline 
	\end{tabular}
	\label{tab:lambda}
\end{table}

In Table \ref{tab:ablation}, \# 1 has only the normal sequence branch of  the  cTransformer. 
That is, the structure of \# 1 in Table  \ref{tab:ablation} is equivalent to  
Transformer \cite{Transformer}.
Conversely, \# 2 has only the reverse sequence branch. 
Except for the result of \# 2 from the reverse sequence branch, the rest of the results are predicted by the normal sequence branch.
The model in \# 4 outperforms \# 3 without context loss, 
indicating that the context loss is beneficial to the model to effectively integrate the forward information related to the target in the normal sequence and the backward information related to the target in the reverse sequence.
With the support of contextual information, the model can more accurately identify and effectively confirm the target event.

Table \ref{tab:lambda} further controls the scale of different losses in a fine-grained manner to filter out the optimal combination of coefficients.
 Table 3 attempts to control variables to compare the performance of models with different combinations of coefficients. 
Finally, giving the same weight to $\mathcal{L}_{normal}$ and $\mathcal{L}_{context}$, and lightening the weight of $\mathcal{L}_{reverse}$ achieves the best AUC in \# 4. 
This reveals that in the experiments, the cTransformer should focus on capturing the forward and contextual information, while putting the backward information in a secondary position for better event sequence recognition.

After the structure of the proposed context Transformer and hyperparameters of losses are determined, 
Table \ref{tab:baselines} compares the cTransformer with Baseline and other methods related to audio event sequence analysis. 
To analyze the recognition ability of different models to polyphonic audio events from multiple perspectives, 
several metrics are adopted to evaluate the AT results of models in Table \ref{tab:baselines}, while the classical \textsl{BLEU} is still used for SAT.
In Table \ref{tab:baselines}, BLSTM-CTC \cite{wangyun}, which uses only LSTM to extract acoustic representations to identify polyphonic audio event sequences, has the worst overall performance.
The CBGRU-CTC \cite{csps_ctc} with a composite convolutional recurrent neural network outperforms the BLSTM-CTC overall, which implies the superior ability of the convolutional layer in feature extraction.
CGLU-BGRU-CTC \cite{dcase2018_ctc} and CBGRU-GLU-CTC \cite{hou2019sound} with 
GLU
assembled in convolutional layers and both convolutional and recurrent layers, respectively, do not perform very well overall, although they outperform CBGRU-CTC in some metrics.
 This paper also shows the results of the default Transformer \cite{Transformer} with 6-layer encoder and decoder. Possibly due to the size of the polyphonic audio dataset containing diverse and complex event sequences is not large, the performance of Transformer is close to that of the CTC-based methods.
Overall, the cTransformer achieves better results in both AT and SAT.
Since data augmentation is not used in the previous CTC-based methods, 
none of the above models have trained with data augmentation for a fair comparison.

\begin{table}[b]\footnotesize
	\setlength{\abovecaptionskip}{0cm}   
	\setlength{\belowcaptionskip}{-0.5cm}   
	\renewcommand\tabcolsep{1.5pt} 
	\centering
	\caption{Comparison of SAT and AT results with other methods related to the analysis of audio event sequences.}
	\begin{tabular}{
	p{2.835cm}<{\centering}|
	p{0.68cm}<{\centering}
	p{0.72cm}<{\centering}
	p{0.68cm}<{\centering}
	p{0.99cm}<{\centering}
	p{0.6cm}<{\centering}|
	p{0.8cm}<{\centering}} 
		\hline 
	\multirow{2}{*}{\makecell[c]{\textsl{Method}}}	& \multicolumn{5}{c|}{\textsl{AT}} & \textsl{SAT}\\
		\cline{2-7} 
		\specialrule{0em}{0.07em}{0.pt}
		 & \textsl{P (\%)} & \textsl{R (\%)} & \textsl{F (\%)} & \textsl{Acc (\%)} & \textsl{AUC} & \textsl{BLEU}\\
		\hline 
		\specialrule{0em}{0.07em}{0.pt}
		\textsl{BLSTM-CTC} \cite{wangyun} & 69.73 & 50.12 & 58.32 & 89.47 & 0.713 & 0.323 \\
		
		\textsl{CBGRU-CTC} \cite{csps_ctc} & 67.79 & 63.39 & 63.23 & 90.93 & 0.793 & 0.475 \\
		
		\textsl{CGLU-BGRU-CTC} \cite{dcase2018_ctc} & \textbf{79.87} & 60.99 & 69.17 & 90.48 & 0.786 & 0.468 \\
		
		\textsl{CBGRU-GLU-CTC} \cite{hou2019sound} & 75.97 & 64.30 & 69.65 & 91.77 & 0.787 & 0.463 \\
		
		\textsl{Transformer} \cite{Transformer} & 67.24  & 64.53 & 65.86 & 90.17 & 0.785 & 0.432 \\
		
		\textsl{cTransformer} & 75.66 & \textbf{67.61} & \textbf{71.41} & \textbf{92.05} & \textbf{0.805} & \textbf{0.505} \\
		\hline 
	\end{tabular}
	\label{tab:baselines}
\end{table}

 \label{ssec:figure-att}
\begin{figure}[t]
	\setlength{\abovecaptionskip}{0.1cm}  
	\setlength{\belowcaptionskip}{-0.55cm}   
	\centerline{\includegraphics[width = 0.45 \textwidth]{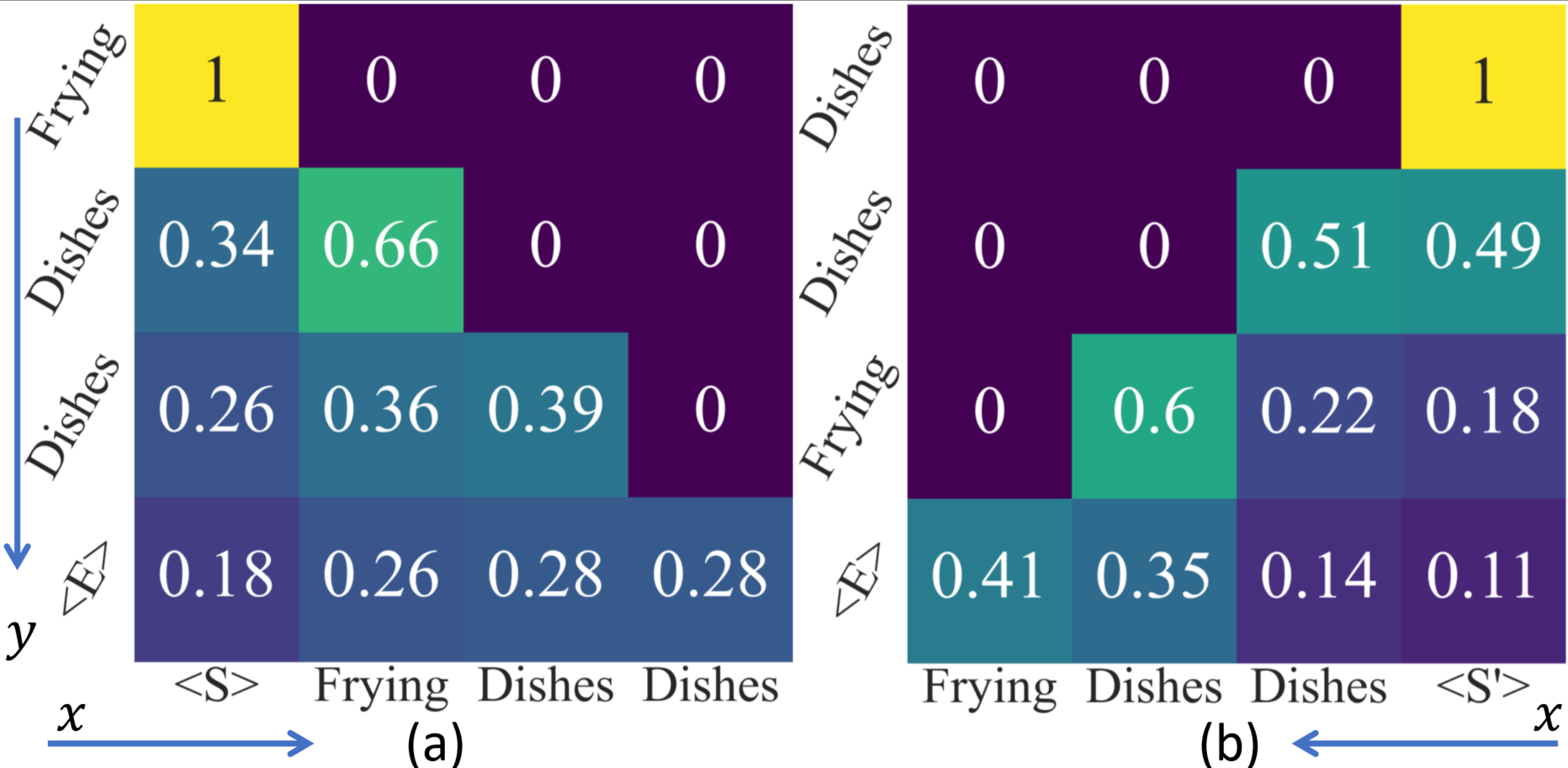}}
	\caption{Attention score from the masked MHA in decoder. 
	Subgraph (a) and (b) are from the normal and reverse sequence branches, respectively.
	The x-axis is each event predicted by the autoregressive way, y-axis is the corresponding reference event.
	}
	\label{att}
\end{figure}

To gain a more intuitive insight of the performance of the model on polyphonic audio event sequences, for the same audio clip, Figure~\ref{att} shows the distribution of attention scores from masked MHA of the normal  and reserve sequence branches.
In Figure~\ref{att} (a), 
after inputting 
\textit{\footnotesize \textless{S}\textgreater}
the attention value for \textit{\footnotesize \textless{S}\textgreater} is 1,
then combining 
acoustic representations from the encoder, 
the model predicts the next event should be 
\textit{\footnotesize frying}
(the event corresponding to the 2nd column of x-axis), and the reference event label is \textit{\footnotesize frying}
(the event corresponding to the 1st row of y-axis). 
Then, the input is
\textit{\footnotesize ``\textless{S}\textgreater, frying''}
attention values for the two events are 0.34 and 0.66, respectively, 
and the next event is predicted to be 
\textit{\footnotesize dishes}
(the event corresponding to the 3rd column of x-axis), and reference event label is \textit{\footnotesize dishes}
(the event corresponding to the 2nd row of y-axis). 
Finally, when the input is 
\textit{\footnotesize ``\textless{S}\textgreater, frying, dished, dishes''}, 
based on acoustic representations, the model judges that the event sequence is complete, and subsequently outputs 
\textit{\footnotesize \textless{E}\textgreater}
(the event corresponding to the 4th row of y-axis) to indicate the inference stops.
After the autoregressive process,  the predicted event sequence 
$p=$\textit{\footnotesize ``frying, dished, dishes''}
is obtained,
the reference label sequence $y$ is 
\textit{\footnotesize ``frying, dished, dishes''}.
The exact match between $p$ and $y$ indicates that the cTransformer successfully fuses frame-level acoustic representations from the encoder with clip-level event cues from the decoder to jointly infer the event sequence.
In Figure~\ref{att} (b), the attention scores from reverse sequence branch for the same audio clip are different from attention scores for forward inference in Figure~\ref{att} (a).
Guided by 
\textit{\footnotesize $\textless{S'}\textgreater$},
the reverse sequence branch combing audio representations successfully predicts the reverse event sequence
$p\prime=$\textit{\footnotesize ``dished, dishes, frying''},
the corresponding label $y\prime$ is 
\textit{\footnotesize ``dished, dishes, frying''}.
The match of $p\prime$ and $y\prime$ indicates that with the assistance of different prediction cues and mask matrices, the cTransformer effectively infers the event sequence from normal and reverse directions,  which implies that the model is effective for modeling contextual information.




\vspace{-0.2cm}
\section{Conclusions}

This paper first introduces Transformer into SAT. 
To utilize the context information of audio event sequences, cTransformer
is proposed to recognize diverse event sequences in polyphonic audio clips.
The cTransformer can automatically assign different attention scores to the existing information to effectively model contextual information and accurately infer the event, 
then frame-level acoustic representations and clip-level event cues are efficiently fused to successfully identify and predict event sequences implicit in audio clips.
Future work will explore the performance of cTransformer using fully bidirectional information to infer audio event sequences on more datasets.

\vfill\pagebreak
\bibliographystyle{IEEEtran}

\bibliography{mybib}


\end{document}